\documentstyle[psfig,preprint,graphicx,tighten,floats,aps]{revtex}
\makeatletter

\def\ut#1{\rlap{\lower1ex\hbox{$\sim$}}{#1}}

\def\S{\Sigma}
\def\d{{\rm d}}

\def\ba{\begin{eqnarray}}
\def\ea{\end{eqnarray}}
\def\be{\begin{equation}}
\def\ee{\end{equation}}
\def\={\mathrel{\widehat\mathalpha{=}}}
\def\puto#1{\rlap{\raise.5ex\hbox{\char'27}}{#1}}
\def\ADM{{\rm ADM}}
\def\BH{{\rm hor}}
\def\sol{{\rm sol}}
\def\t{{(t)}}
\def\b{{\rm binding}}
\def\a{{\rm avail}}
\def\i{{\rm initial}} 
\def\f{{\rm final}}
\def\I{{\cal I}^+}
\def\in{{\rm inter}}

\preprint{\vbox{\baselineskip=12pt
\rightline{CGPG-00/11-2}
\rightline{ICN-UNAM-00/14}
\rightline{gr-qc/0011081}
}}

\begin{document}
\title{Hairy Black Holes, Horizon Mass and Solitons} 
\author {Abhay Ashtekar${}^1$\thanks{E-mail: ashtekar@gravity.phys.psu.edu},
Alejandro\ Corichi${}^2$\thanks{E-mail: corichi@nuclecu.unam.mx}, and Daniel 
Sudarsky${}^2$\thanks{E-mail: sudarsky@nuclecu.unam.mx} } 
\address{{1.} Center for Gravitational Physics and Geometry\\
Department of Physics, Penn State, University Park, PA 16802, U.S.A.}
\address{{2.} Instituto de Ciencias Nucleares\\
Universidad Nacional Aut\'onoma de M\'exico,
A. Postal 70-543, M\'exico D.F. 04510, M\'exico.}

\maketitle

\begin{abstract}
Properties of the horizon mass of hairy black holes are discussed with
emphasis on certain subtle and initially unexpected features.  A key
property suggests that hairy black holes may be regarded as `bound
states' of ordinary black holes (without hair) and colored solitons.
This model is then used to predict the qualitative behavior of the
horizon properties of hairy black holes, to provide a physical
`explanation' of their instability and to put qualitative constraints
on the end point configurations that result from this instability.
The available numerical calculations support these predictions.
Furthermore, the physical arguments are robust and should be
applicable also in more complicated situations where detailed
numerical work is yet to be carried out.

\end{abstract}

\section{Introduction}
\label{s1}

\begin{figure}
\centerline{
\hbox{\psfig{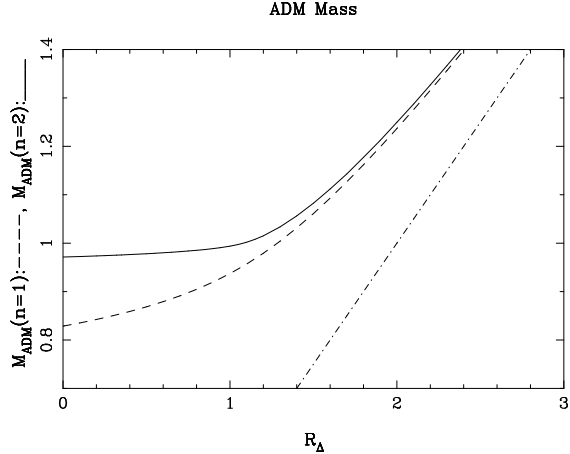}}}
\bigskip
\caption{The ADM mass as a function of the horizon radius $R_{\Delta}$
of static spherically symmetric solutions to the Einstein-Yang-Mills
system (in units provided by the Yang-Mills coupling constant).
Numerical plots for the colorless, $n=0$, and the first two families,
$n =1,2$ of colored black holes are shown. (Note that the $y$-axis
begins at $M = 0.7$ rather than $M= 0$.) }
\label{fig1}
\end{figure}

In the Einstein-Maxwell-Klein-Gordon theory, static black holes are
remarkably simple objects.  They are completely characterized by their
mass and (electric and/or magnetic) charge, \textit{evaluated at
infinity}. (For reviews, see \cite{pc,mh}).  As is well-known, this
simplicity does not persist once non-Abelian gauge fields are
incorporated \cite{pb1,pb2,nqs}.  For example, for a sufficiently
large value of the ADM mass $M_{\ADM}$, the Einstein-Yang-Mills theory
admits distinct spherical, static black hole solutions, labelled by an
integer $n$, all with zero charge at infinity%
\footnote{An infinite family of static spherical solutions was
shown to exist in \cite{Smoller:1993bs,Breitenlohner:1994es}.
Furthermore, unlike in the Einstein-Maxwell theory, there
also exist \textit{non-spherical} static black hole solutions.
Although one does not yet have a full control on the entire static
sector, the known family of static solutions with zero charge at
infinity can labelled by the horizon area and \textit{two} integers
\cite{kk}.}
(see Fig.~\ref{fig1}). Although in terms of conserved quantities at
infinity these solutions are indistinguishable from the Schwarzschild
black hole of mass $M_{\ADM}$, in the interior their geometries and
Yang-Mills fields are quite different from one another.  Even the
horizon structure varies as we pass from one $n$ to another; for
instance, for fixed $M_{\ADM}$, the horizon area is a decreasing
function of $n$.  Thus, the rigidity tying conserved quantities at
infinity to the structure in the interior does not extend to the
non-Abelian context.  Following the common terminology, if this
failure occurs we will say that the theory admits `hair' and refer to
these solutions as `hairy black holes'.

Of central importance to this paper is the notion of the horizon (or
black hole) mass $M_{\BH}$ in general (Einstein-matter fields)
theories with hair.  In the Einstein-Maxwell theory, for static black
hole solutions one can set $M_{\BH} = M_{\ADM}$.  However, outside
these \textit{globally} time-independent situations, this
identification is not viable.  For, there may be gravitational
radiation in a region far removed from the black hole, and hence
irrelevant for any reasonable notion of the horizon mass $M_{\BH}$,
which would nonetheless obviously contribute to $M_{\ADM}$. Since
hairy black holes are known to be unstable, such dynamical situations
arise naturally in their study. Furthermore, as discussed in Section
\ref{s3.1}, \textit{even in the static context}, $M_{\ADM}$ is a poor
measure of the mass of these black holes.  Therefore, it is desirable
to have a general, systematic approach to the problem of defining
horizon mass. In this paper, we will reach this goal by using the
recent \textit{isolated horizon} framework, based on Hamiltonian
methods \cite{abf,ac:ds,afk}.

Closely related to hairy black holes are the solitonic solutions to
Einstein-Yang-Mills equations
\cite{bk,Smoller:1993,Breitenlohner:1994es}. Indeed, it was the
Bartnik-McKinnon \cite{bk} discovery of these solitons 
that shook the then prevailing view that there is no qualitative
difference between Einstein-Maxwell and Einstein-Yang-Mills theories
and paved the way for discovery of hairy black holes. The purpose of
this paper is to use the isolated horizon framework to gain new,
qualitative insights in to the relation between colored black holes
and their solitonic analogs and to point out certain subtleties that
arise as we move away from the Einstein-Maxwell context and admit
increasing complexity in the hair.  A key property of the horizon mass
suggests that a colored black hole can be regarded as a `bound state'
of an ordinary, uncolored black hole and a colored soliton.  This
physical picture turns out to be very useful.  It enables one to
regard the total (ADM) mass as a sum of three terms: the mass of a
bare (i.e. uncolored) black hole, the mass of the appropriate soliton
and the gravitational binding energy between the two.  Physically
expected properties of the gravitational binding energy (such as its
negative sign) then determine the qualitative behavior of the horizon
attributes of colored black holes (such as surface gravity).  These
predictions of the model have already been borne out by numerical
calculations in simple cases.  We will also use the model to argue
that instability of colored black holes `stems' from that of solitons
and constrain the final outcome of small perturbations.  Finally, as
is well-known, quasi-local definitions of mass in general relativity
tend not to possess all the properties one is accustomed to from field
theories in flat space.  We will see that while the isolated horizon
framework provides an internally consistent viewpoint and strategy,
this very consistency implies that the resulting horizon mass has
certain counter-intuitive features. They turn out to have important
physical implications, in particular to the stability properties of
magnetically charged Reissner-Nordstr\"om solutions.

The transition from Abelian to non-Abelian gauge fields is accompanied
by the emergence of new scales with dimensions of energy (or length).
The classical Einstein-Maxwell theory has no such scale.  The coupling
constant $g$ of the Einstein-Yang-Mills theory provides an energy
scale $M_0= 1/(g\sqrt{G})$ (in the c=1 units) which signals the onset
of new branches (labelled by $n$) of spherical static solutions (see
Fig.~\ref{fig1}). In the Einstein-Yang-Mills-Higgs theory, the mass
and the self-coupling constant of the Higgs field provide additional
scales.  Typically, as the number of such scales increases, the
departure from the Einstein-Maxwell behavior becomes more significant
and the number of unexpected, subtle features grows.

In Section~\ref{s2} we recall the definition of the horizon mass in
the Einstein-Maxwell theory.  In Section~\ref{s3} we consider
Einstein-Yang-Mills black holes, discuss a key relation between the
horizon mass of static, hairy black holes and the mass of the
corresponding solitons, introduce the heuristic model of colored black
holes mentioned above and derive some of its predictions. In
particular, we will show that the notion of the horizon mass can be
used to `explain' why the magnetically charged Reissner-Nordstr\"om
solutions are unstable in the Einstein-Yang-Mills theory
\cite{stability} although they are stable in the Einstein-Maxwell
theory \cite{moncrief}.  
For definiteness, in this section we will restrict ourselves
to Einstein-Yang-Mills theory. However, modulo certain subtleties,
most considerations of this section will apply also to hairy black
holes in more general theories. In Section~\ref{s4} we consider such
theories and discuss the relevant subtleties.  These arise from the
fact that, while in the Einstein-Yang-Mills theory  hairy black holes
belong to {\it disjoint} families carrying discrete labels (e.g.,
integers), in more general theories these families intersect. 
Section~\ref{s5} provides a brief summary.

\section{Isolated Horizons and horizon mass\\
in Einstein-Maxwell theory}
\label{s2}

To define the horizon mass, it suffices to consider the so-called
`weakly isolated horizons'.  The precise definition and a detailed
discussion can be found in \cite{afk} but will not be needed here.  In
essence, a weakly isolated horizon $\Delta$ is an expansion-free,
null, 3-dimensional sub-manifold $\Delta$ of the 4-dimensional
space-time $(M, g_{ab})$, equipped with an equivalence class of
future-directed null normals $\ell^a$ which Lie-drag (certain
components of) the intrinsic connection on $\Delta$.%
\footnote{Here, $\ell^a$ is equivalent to $\tilde{\ell}^a$ if an only
if $\tilde{\ell}^a = c \ell^a$ where $c$ is a positive constant on
$\Delta$.  Because the Maxwell stress-energy condition has the
property that $-T_{ab}\ell^b$ is future-directed, the Raychaudhuri
equation equation implies that the requirement that $(\Delta, \ell^a)$
be expansion-free is equivalent to the requirement that $\ell^a$
Lie-drags the intrinsic (degenerate) metric on $\Delta$.}
The notion extracts from the definition of a Killing horizon just
those properties \textit{local to the horizon} that are needed to
obtain a consistent Hamiltonian framework, define the horizon energy,
angular momentum and other charges, and prove the zeroth and first
laws \cite{afk,abl2}.  In the non-rotating case now under
consideration, the vector field $\ell^a$ on $\Delta$ plays the role of
the static Killing field.  However, as explicit examples and general
constructions show, the space-time need not admit a Killing field in
\textit{any} neighborhood of $\Delta$ \cite{jl,afk}.

To define the horizon mass $M_\Delta$, one proceeds as follows.
Recall that, in the standard asymptotically flat context, the ADM
energy $E^t_{\ADM}$ arises as the boundary term at spatial infinity in
the expression of the Hamiltonian generating evolution along an
asymptotic time-translation $t^a$.  Now the idea is to use as phase
space the sector of general relativity consisting of space-times which
are asymptotically flat at spatial infinity \textit{and} admit an
isolated horizon $\Delta$ as \textit{internal boundary}.  The
Hamiltonian generating an appropriate time-translation then has two
surface terms, one at infinity, $E^\t_\infty$, and one at the internal
horizon boundary, $E^\t_\Delta$.  As one might expect, the first is
precisely the ADM energy $E^t_\infty = E^t_{\ADM}$.  The second,
$E^\t_\Delta$, is interpreted as the horizon energy.

However, in the detailed implementation of this strategy a subtlety
arises which will play an important role in this paper.  At infinity,
all metrics in the sector under consideration approach a fixed
Minkowskian metric $\eta_{ab}$.  Therefore, we can simply fix a
time-translation Killing field of $\eta_{ab}$ and demand that $t^a$
agree with that vector field in a neighborhood of infinity.  At the
internal boundary, by contrast, the geometry is not universal.
Therefore, given a candidate time-translation near $\Delta$ for one
space-time in our sector, it is not a priori clear how to choose the
\textit{same} time translation in \textit{another} space-time (also
belonging to the sector).  To compute the Hamiltonian, one must fix
this ambiguity, i.e., choose appropriately a vector field $t^a$ for
each space-time in the phase space.  In the non-rotating case, it is
natural to set $t^a= c\ell^a$ on $\Delta$.  However, while $c$ is
constant on $\Delta$ in any one space-time, it may vary from one
space-time to another, i.e., can be a function (of the horizon
parameters) on the full phase space.  Put differently, in the
numerical relativity terminology, one is led to allow \textit{live}
evolution vector fields $t^a$ (or, equivalently, lapse-shift pairs
that depend on the point on phase space) on space-time.

Let us fix such a live vector field $t^a$.  The evolution along it
defines a vector field $X_t$ on the whole phase space.  It is then
natural to ask if this evolution is Hamiltonian, i.e., if it preserves
the symplectic structure.  Somewhat surprisingly, the answer is in the
affirmative \textit{if and only if} there exists a function
$E^\t_\Delta$ of the horizon area $a_\Delta$ and electric charge
$Q_\Delta$ such that
\be \label{1law}
\delta E^\t_\Delta \= \frac{\kappa_\t}{8\pi G}\,\, \delta  a_\Delta + 
\Phi_\t \delta Q_\Delta\, .  \ee
for arbitrary tangent vectors $\delta$ to the phase space \cite{afk}.
Here, $\=$ stands for equality restricted to $\Delta$, the surface
gravity $\kappa_\t$ is defined by $t^a$ via $t^a\nabla_a t^b \=
\kappa_\t t^b$, and $\Phi_\t$ is the electric potential defined by
$t^a$ on $\Delta$.  ($\kappa_\t$ is called `surface gravity' because
$t^a \= c\ell^a$, with $c$ a constant on $\Delta$. The potential
$\Phi_\t$ is unimportant for the purposes of this paper.)  Note that
(\ref{1law}) implies that $\kappa_\t$ and $\Phi_\t$ are also functions
only of $a_\Delta$ and $Q_\Delta$; they can not depend on other
properties of the solution under consideration.

If $E^\t_\Delta$ is interpreted as the horizon energy, (\ref{1law}) is
precisely the first law of black hole mechanics!  Thus, the necessary
and sufficient condition for the evolution generated by a live vector
field $t^a$ to be Hamiltonian is precisely that the first law hold.
A live vector field satisfying this condition is said to be
\textit{permissible}.  The Hamiltonian $H^\t$ generating evolution
along a permissible vector field $t^a$ is given by
\be \label{ham} H^\t = \int_\S ({\rm constraints})\, \d^3x + E^\t_{\ADM} -
E^\t_{\Delta} \ee
where $\S$ is any partial Cauchy slice extending from the isolated horizon 
$\Delta$ to spatial infinity.

One can now ask if such live vector fields exist on the whole phase 
space.%
\footnote{Since the volume term in the expression (\ref{ham}) is a
linear combination of constraints, only the boundary values of the
live evolution vector field $t^a$ matter on `on shell'.  Therefore,
while discussing evolution vector fields, we will be concerned only
with their boundary values.}
The answer is in the affirmative \cite{afk}.  Choose any (regular)
function $\kappa_0$ of $a_\Delta$ and $Q_\Delta$.  Then, one can show
that there exists a permissible live vector field $t^a$ with
$\kappa_\t = \kappa_0$.  Thus, in fact, there exists an infinite
family of permissible live vector fields, each defining a horizon
energy $E^\t_\Delta$ and leading to a first law.  While the phase
space meaning $E^\t_\Delta$ is transparent, for a general permissible
vector field $t^a$, its space-time significance is not so clear.  Is
there perhaps a canonical choice for the boundary value of $t^a$ on
$\Delta$?  In the Einstein-Maxwell theory, the answer is in the
affirmative.  Let us demand that the required live vector field
$t_0^a$ be such that in \textit{each} static
(i.e. Reissner-Nordstr\"om) solution $t_0^a$ coincides with that
static Killing field which is unit at infinity.  This is a natural
choice.  Furthermore, it fixes $\kappa{(t_0)}$ uniquely as a function
of $a_\Delta$ and $Q_\Delta$.  Therefore, although we fixed the
normalization of $t_0^a$ \textit{only} on the static solution, since
there exists a unique static black hole solution for each ($a_\Delta$,
$Q_\Delta$), the functional form of $\kappa_{(t_0)}$ is completely
fixed on the whole phase space.  This in turn selects the boundary
value of $t_0^a$ on $\Delta$ and determines $E^{(t_0)}_{\Delta}$
everywhere on the phase space.

It is then natural to set the black hole mass
\be M_{\BH} =  E^{(t_0)}_{\Delta} \ee
for any space-time in the phase space.  This definition of $M_{\BH}$
has several pleasing features.  First, because $t_0$ is permissible,
(in an obvious notation) the corresponding first law (\ref{1law})
takes the familiar form 
\be
\delta M_{\BH} = \left({\kappa \over 8\pi G}\right) \,\delta a_\Delta 
+\Phi \delta Q_\Delta\, ,\ee
but now on the \textit{entire} phase space, including non-static
space-times admitting gravitational and electromagnetic radiation.
Second, if the evolution vector field coincides with a static Killing
field, general phase space considerations imply that the total
Hamiltonian is constant on each connected component of the static
sector.  Since in the Einstein-Maxwell theory, there is no natural
energy scale, this constant is necessarily zero.  Eq (\ref{ham}) now
implies that on static solutions, $M_{\BH} = M_{\ADM}$.  Finally, in
presence of radiation, assuming that the horizon extends all the way
to future time-like infinity $i^+$ and that the structure of $i^+$ is
`the same as that in Reissner-Nordst\"orm space-times', one can show
\cite{abf},
\be  M_{\ADM} - M_{\BH} = E_{\I}\, ,\ee
the total energy radiated across future null infinity, ${\I}$.
(Equivalently, even in presence of radiation, $M_{\BH}$ is the future
limit of the Bondi energy as one approaches $i^+$ along ${\cal I}^+$.)
Thus, in the Einstein-Maxwell theory, the horizon mass $M_{\BH}$ is
defined everywhere on the phase space and has the properties one
intuitively expects.

Let us summarize.  At infinity, $M_{\ADM}$ is the norm of the
asymptotic 4-momentum.  Because the horizon is in a strong field
region, we do not have a 4-dimensional translation group on $\Delta$,
whence there is no well-defined notion of horizon 4-momentum.
Therefore, we need an alternate extra structure to define $M_{\BH}$.
The idea is to use a preferred time-translation $t^a_0$ and identify
$M_{\BH}$ with the horizon energy $E^{\t_0}_\Delta$ defined by this
translation.  This task could be carried out because, thanks to the
no-hair theorems, we could choose a preferred, live evolution field
$t_0^a$ for \textit{all} space-times in the full phase space and use
it to introduce a canonical definition of $M_{\BH}$ on the whole phase
space.

\section{Yang-Mills hair, horizon mass and stability}
\label{s3}

This section is divided in to four parts.  In the first, we recall the
known results from \cite{ac:ds,afk}.  In the second, we introduce a
model of colored black holes as bound states of the ordinary black
holes and appropriate solitons and show that it predicts all the
qualitative features of the horizon structure of colored black holes
in equilibrium.  In the third part, we use the model to discuss
instability of colored black holes and put constraints on the final
configurations resulting from small perturbations.  In the fourth
part, we focus on the `embedded Abelian black holes' and show that the
horizon mass of these black holes is higher if they are regarded as
Yang-Mills black holes than it is if they are regarded as
Einstein-Maxwell black holes.  Thus, because the horizon-mass is a
phase-space rather than space-time notion, in this case it turns out
to be `theory dependent'.  This is a striking and, at a first glance,
surprising property.  It has an important physical consequence,
briefly noted in \cite{ac:ds}: using considerations of the first three
sub-sections, we will show in detail that it `explains' why these
solutions are unstable in the non-Abelian sector although they are
stable in the Einstein-Maxwell theory.  This is another striking
example of the usefulness of the notion of horizon mass.

\subsection{Horizon mass of colored black holes}
\label{s3.1}

Let us now consider the Einstein-Yang-Mills theory with gauge group
$SU(2)$.  In this case the Reissner-Nordstr\"om solutions provide a
2-parameter family of static, spherically symmetric, Abelian black
holes, labelled by $(M_{\ADM}, Q)$, where $Q$ is now associated with a
fixed $U(1)$ subgroup of $SU(2)$.  In the Maxwell case, if the
potential $A_a$ is a connection 1-form on a trivial bundle, the
magnetic charge vanishes identically (and, on non-trivial bundles, it
is quantized).  In the Yang-Mills case, by contrast, there is a
1-parameter family of magnetically charged, static solutions (with a
fixed magnetic charge $P_0$) \textit{even on the trivial $SU(2)$
bundle} \cite{review}.  These are called `embedded Abelian solutions'
because they are isometric to a family of magnetically charged
Reissner-Nordstr\"om solutions and the isometry maps the Maxwell field
strength to the Yang-Mills field strength.  The only difference is in
the form of the connection.  We will return to these black holes in
the subsection \ref{s3.4}.  Of direct interest to this subsection are
the families of `genuinely' non-Abelian, static solutions. Each family
is parametrized by one continuous parameter ($a_\Delta$).  An infinite
number of families, labelled by two integers, is known to exist of
which spherically symmetric solutions constitute a sub-class labeled
by one integer of Fig.~\ref{fig1} 
\cite{Smoller:1993bs,Breitenlohner:1994es}.

We will now show that, in the Einstein-Yang-Mills theory, the ADM mass
fails to be a good measure of the black hole mass \textit{even in the
static sector}.  Consider, for instance, the branch of spherical,
static black holes labelled by $n \not= 0$ (Fig.~\ref{fig1}).  Let us
decrease the horizon area along this branch.  In the zero area limit,
the solution is known to converge point-wise to a regular, static,
spherical solution, representing a Einstein-Yang-Mills
\textit{soliton} \cite{bk,review}.  This solution has, of course, a
non-zero ADM mass $M_{\ADM}^{\sol}$, which equals the limiting value
of $M_{\ADM}^{\rm BH}$.  However, in this limit, there is no black
hole \textit{at all}!  Hence, this limiting value of the ADM mass can
not be meaningfully identified with any horizon mass.  By continuity,
then, $M_{\ADM}^{\rm BH}$ can not be taken as an accurate measure of
the horizon mass for any black hole along any $n\not= 0$ branch.

Can one use the isolated horizon framework to define $M_{\BH}$ of
static black holes in the $n\not= 0$ families?  The answer is in the
affirmative but there is a subtlety.  Considerations of Section
\ref{s2} leading to eqs (\ref{1law}) and (\ref{ham}) go through also
in this case \cite{afk}.  Permissible vector fields can be constructed
as in Section \ref{s2} and, for each permissible vector field $t^a$,
we again have a well-defined notion of the horizon energy
$E^\t_{\Delta}$ on the entire phase space, and the associated first
law.  However, since the no hair theorem now fails, we can no longer
single out a \textit{preferred} live vector field $t^a_0$ on the
\textit{full} phase space by using static solutions and set $M_{\BH} =
E^{t_0}_{\BH}$.  We could focus just on the electrically charged,
static Abelian solutions.  Surface gravity $\kappa^{\rm Abel}$ of the
properly normalized Killing field of these solutions does provide us,
as in the Maxwell case, with a specific function $\kappa^{\rm Abel}
(a_\Delta, Q_\Delta)$ and we can again use it to obtain a preferred,
permissible, live vector field $t^a_{\rm Abel}$ and corresponding
horizon energy $E^{\rm Abel}_{\Delta}$.  For these static solutions,
we can again set $M_{\BH} = E^{\rm Abel}_\Delta$.  However,
$\kappa^{\rm Abel}$ fails to agree with the surface gravity of the
properly normalized Killing field in hairy static solutions.  Hence
there is no reason to interpret $E^{\rm Abel}_\Delta$ as the horizon
mass of the hairy solutions.  To summarize, because the uniqueness
theorem fails ---i.e., because of the hair--- we can not single out a
canonical function $\kappa_0$ of the horizon parameters which agrees
with the surface gravity of the (properly normalized) Killing field on
\textit{all} static solutions.  Therefore, the Einstein-Maxwell
prescription for defining the horizon mass $M_{\BH}$ on the
\textit{full} phase space does not extend to theories with hair.

However, there \textit{does} exist a natural strategy to assign a
horizon mass to any \textit{static} solution \cite{afk}.  For
definiteness, let us consider the 1-parameter family of static,
spherical solutions labelled by $n_0$.  (However, our discussion and
conclusions apply to other static, hairy black holes as well.)  Since
they have zero electric charge, along this branch the surface gravity
(defined by the properly normalized Killing field) is a function
$\kappa_{(n_0)}(a_\Delta)$ only of the area $a_\Delta$.  By setting
$\kappa_0 = \kappa_{(n_0)}$ in the construction outlined in the last
sub-section, we obtain a permissible, live vector field $t^a_{n_0}$ on
the entire phase space. The corresponding horizon energy
$E^{(n_0)}_\Delta$ is given by \cite{ac:ds,afk}:
\be \label{energy}
E^{(n_0)}_\Delta = \frac{1}{2G}\int_0^{R_\Delta} \beta_{(n_0)}(r)\, \d r 
\, ,\ee
where $R_\Delta$ is the horizon radius (i.e., the horizon area is
given by $a_\Delta = 4\pi R_\Delta^2$),\,\, and
$\beta_{(n_0)}(R_\Delta) = (2 R_\Delta)\,\,
(\kappa_{(n_0)}(R_\Delta))$.  While this horizon energy
$E_{\Delta}^{(n_0)}$ has no obvious space-time interpretation for
\textit{general} space-times, for static solutions on the branch $n=
n_0$, it is natural to set
\be\label{hormass} 
M^{(n_0)}_{\BH}(R_\Delta) = E^{(n_0)}_\Delta (R_\Delta)\, 
= \frac{1}{2G}\int_0^{R_\Delta} \beta_{(n_0)}(r) \d r\, .
\ee
We will do so.  Note that $M_{\BH}^{(n_0)}(R_\Delta)$ vanishes at
$R_\Delta =0$ (and then monotonically increases with area). Hence this
definition of the horizon mass is free of the problem faced by the ADM
mass, discussed above.

Finally, for any $n$ one can relate the horizon mass $M^{(n)}_{\BH}$
to the ADM mass of static black holes.  Recall first that general
Hamiltonian considerations imply that the total Hamiltonian
(\ref{ham}) is constant on every connected component of static
solutions (provided the evolution vector field $t^a$ agrees with the
static Killing field everywhere on this connected component)
\cite{abf,afk}.  In the Einstein-Maxwell case, there is only one
connected component and there is no energy scale in the theory.
Therefore, we could conclude $H^{(t_0)} = 0$ on all static solutions.
In the Einstein-Yang-Mills case, by contrast, there is an infinite
number of connected components of static solutions \textit{and} there
is a natural energy scale, $M_0 = 1/g\sqrt{G}$.  Therefore, on each
distinct connected component, in principle, the Hamiltonian could be a
\textit{different} multiple of $M_0$.  This is exactly what happens.
Since the Hamiltonian is constant on any $n$-branch, we can evaluate
it at the solution with zero horizon area.  This is just the soliton,
for which the horizon area $a_\Delta$, and the horizon mass $M_\BH$
vanish.  Hence (\ref{ham}) implies $H^{(t_0,n)} = M_{\sol}^{(n)}$.
Thus, we conclude:
\be \label{ymmass}
M^{(n)}_{\sol} = M^{(n)}_{\ADM} - M^{(n)}_{\BH} \ee
on the entire $n$th branch \cite{ac:ds,afk}.  Thus, the ADM mass
contains two contributions, one attributed to the black hole horizon
and the other to the outside `hair', captured by the `solitonic
residue'.

\subsection{A physical model of colored black holes}
\label{s3.2}

Considerations of Section \ref{s3.1} lead to a heuristic, but quite
powerful physical model of a colored black hole.  In this subsection
we will develop this model and derive some of its consequences on the
equilibrium properties of static black holes.

Recall that for static black holes in the Einstein-Maxwell theory, the
future limit $M_{\rm Bondi}^{i^+}$ of the Bondi mass along ${\cal
I}^{+}$ equals the horizon mass $M_{\BH}$.  What is the situation for
the non-Abelian black holes of the Einstein-Yang-Mills theory?  In the
zero area limit along any $n\not= 0$ branch, one is left with a
soliton, i.e., with a regular bound state which persists all the way
to $i^+$ and thus contributes to the mass at $i^+$.  Because the
solution is static, the ADM mass, the Bondi mass and the mass at $i^+$
are all equal.  As we move away from zero area, a true black hole
appears.  For small areas, (\ref{hormass}) implies that
$M_{\BH}^{(n_0)}$ is small.  But, in relation, $M_{\ADM}^{(n_0)}$ can
be large because `there is still a solitonic contribution to the ADM
mass', quantified in (\ref{ymmass}).  In the Einstein-Maxwell case,
there are no solitons; there is no `residue' at $i^+$ apart from the
black hole itself.  This is why for static black holes, $M_{\BH}$
equals $M_{\ADM}$ and charges at infinity suffice to characterize the
fields on the horizon.  In the Einstein-Yang-Mills case, by contrast,
the conserved quantities at infinity do \textit{not} in general
capture the whole story.  For hairy black holes, the structure at
${\cal I}^+$ (or $i^0$) does not rigidly determine the structure at
the horizon; there is a solitonic residue at $i^+$ which accounts for
the lack of rigidity.  This suggests that we think of a colored black
hole as \textit{a `bound state' of an ordinary, `bare' black hole and
a `solitonic residue'}.  Indeed, we can trivially rewrite the total
mass of (\ref{ymmass}) as a sum of three parts, one referring to the
soliton, one to the `bare' black hole and one to the binding energy:
\ba \label{binding} 
M_\ADM^{(n)}(R_\Delta) &=& M^{(n)}_{\sol} + 
M^{(0)}_{\BH} (R_\Delta) +(M^{(n)}_{\BH} - M^{(0)}_{\BH})(R_\Delta)
 \nonumber\\
&=& M^{(n)}_{\sol} + M^{(0)}_\BH (R_\Delta) + E_{\b}(R_\Delta)\, ,\ea
where $M^{(0)}_\BH (R_\Delta)= R_\Delta/2G$ is the horizon mass of the
Schwarzschild black hole of radius $R_\Delta$.  Thus, one can say that
the effect of the solitonic residue is to `dress' the black hole mass
from its `bare value' $M^{(0)}_{\BH}(R_\Delta)$ to
$M^{(n)}_{\BH}(R_\Delta)$.  The fact that the total space-time mass
$M_{\ADM}$ is just a sum $M_{\ADM} = M_{\BH}^{(n)} + M_{\sol}^{(n)}$,
without an explicit interaction term, can be understood as follows.
Recall that, in general relativity, the total Lagrangian or the
Hamiltonian is just a sum of the gravitational part without matter and
the matter contribution.  There is no \textit{explicit} interaction
term between matter and gravity; the interaction is `absorbed' in the
description by simply replacing the flat metric in the `bare' matter
terms by the curved, physical metric, making the full expression
covariant.  In an analogous way, the interaction between the soliton
and the black hole is expressed covariantly by simply replacing the
`bare' black hole mass $M_{\BH}^{(0)}(R_\Delta)$ by the `dressed' mass
$M_{\BH}^{(n)}(R_\Delta)$ as in (\ref{ymmass}).

Let us adopt this model and work out some of its implications.  More
precisely, the idea is to use this model to derive qualitative
features of the properties of colored black holes assuming, whenever
necessary, properties of colored solitons.  Since the colored black
holes solutions have been obtained numerically, we do not have simple
analytical expressions of their horizon attributes such as surface
gravity and horizon mass.  Furthermore, typically such quantities have
been computed \textit{only for low values of} $n$.  In the remainder
of this sub-section we will show that all the qualitative features of
the known numerical plots of these quantities as functions of $n$ and
horizon radius $R_\Delta$ can be derived from our model.  Furthermore
our predictions apply to \textit{all} values of $n$ and $R_\Delta$.

First, since $E_{\b}$ is the gravitational binding energy between the
soliton residue and the bare black hole, it must be negative.  Thus,
the model together with the elementary physical fact about the
gravitational binding energy leads us to the first prediction:

\medskip\noindent
i) \textit{$M_{\BH}^{(n)}(R_\Delta) < M_{\BH}^{(0)}(R_\Delta )$ for 
all $n>0$ and all values of the horizon radius $R_\Delta$.}
\medskip

\begin{figure}
\centerline{
\hbox{\psfig{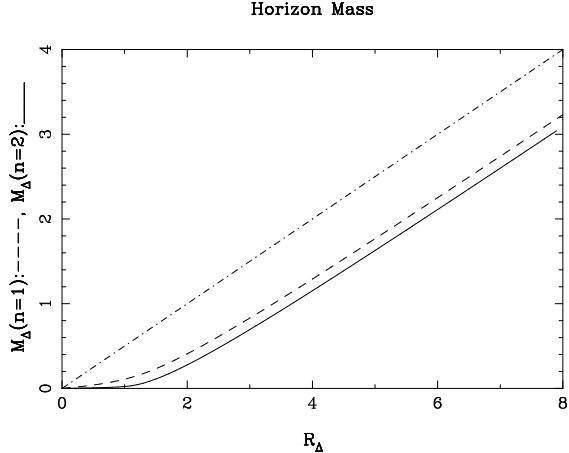}}}
\bigskip
\caption{The horizon mass $M_{\BH}$ as a function of the radius
$R_\Delta$ (in units provided by the Yang-Mills coupling constant).
Numerical plots for $n=0,1,2$ spherical, colored black
holes in the Einstein-Yang-Mills theory are shown.}\label{fig2}
\end{figure}

\noindent At first this conclusion seems somewhat counter-intuitive
since, for a given horizon area, one normally thinks of the colored
black holes as `excited states' and the $n=0$, Schwarzschild black
hole as the `ground state'.  However, that intuition comes from the
properties of the ADM mass: it is known that $M_{\ADM}^{(n)}(R_\Delta)
> M_{\ADM}^{(0)}(R_\Delta)$.  Explicit numerical calculations have in
fact confirmed i) for low values of $n$ (see Fig.~\ref{fig2}): while
$M_{\ADM}^{(n)}(R_\Delta)$ does grow with $n$, $M^{(n)}_{\sol}$ grows
so that i) holds.  The model predicts that this inequality will hold
for all $n>0$.

Next, let us use the expression (\ref{hormass}) of the horizon
mass. From i), we conclude:
$$ \int_{0}^{R_\Delta} \left( \beta_{(n)} - 
\beta_{(0)}\right)(r)\, \d r < 0  $$
\textit{for all} $n$ and $R_\Delta$.  Hence, it follows that
$\left(\beta_{(n)} - \beta_{(0)}\right)(R_\Delta ) <0 $, whence we
conclude\medskip

\noindent
ii) $\left( \kappa_{(n)} - \kappa_{(0)}\right)(R_\Delta) <0 $
\textit{for all} $n>0$ \textit{and} $R_\Delta$.

\begin{figure}
\centerline{
\hbox{\psfig{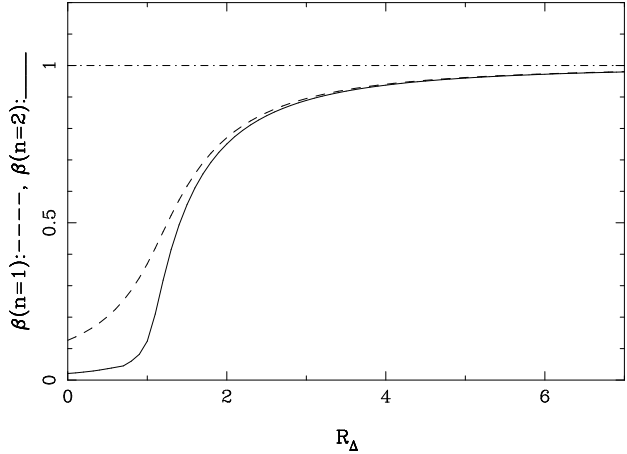}}}
\bigskip
\caption{Numerical plots of $\beta_{(n)}= 2\kappa_{(n)}\, R_\Delta$ as
function of the horizon radius $R_\Delta$ for $n=1,2$ spherical,
colored black holes in the Einstein-Yang-Mills theory. For large
$R_\Delta$, all curves asymptotically approach the $\beta =1$
line.}\label{fig3}
\end{figure}

\medskip
\noindent Again, this prediction is borne out by detailed numerical
simulations for low $n$ (see Fig.~\ref{fig3}).

We can now use the physically expected properties of the gravitational
binding energy to make sharper predictions.  Let us fix $R_\Delta$ and
vary $n$.  Then, the mass $M_{\BH}^{(0)}$ of the bare black hole is
fixed but, as is well-known, the soliton mass $M^{(n)}_{\sol}$
increases with $n$.  Now, since the absolute value of the
gravitational binding energy increases as the mass of either of the two
bound objects grows, $\mid E_{\b}\mid$ must increase with $n$.  This
implies:\medskip

\noindent 
iii) \textit{For any fixed value of $R_\Delta$ both the horizon mass
$M^{(n)}_{\BH}$ and the surface gravity $\kappa_{(n)}$ are 
monotonically decreasing functions of $n$.}

\medskip
\noindent As Fig~\ref{fig2} and Fig~\ref{fig3} show, 
these predictions are borne out numerically for low $n$.

Next, let us consider the complementary situation and keep $n>0$ fixed
and vary $R_\Delta$.  Now the soliton mass $M^{(n)}_{\rm sol}$ is
fixed and the bare black hole mass $M^{(0)}_{\BH}(R_\Delta) =
{R_\Delta/2G }$ increases with $R_\Delta$.  Hence, again (the absolute
value of) the gravitational binding energy must increase with
$R_\Delta$.  Thus, we conclude:\\

\medskip\noindent iv) \textit{For any fixed $n$, $({R_\Delta / 2G} -
M^{(n)}_{\BH})(R_\Delta)$ is a monotonically increasing function of
$R_\Delta$ whence, from (\ref{hormass}),\,\, $\beta_{(n)}(R_\Delta)
<1$ for all $n>0$ and $R_\Delta$}.

\medskip
\noindent Again, Fig.~\ref{fig2} and Fig.~\ref{fig3} show that this
prediction is borne out by numerical studies for low values of $n$.
Furthermore, this property can be directly derived from the definition
(\ref{hormass}) of $M^{(n)}_{\BH}(R_\Delta)$ and prediction ii) above.
Indeed,
$$ \left({R_\Delta \over 2G} - M^{(n)}_{\BH}\right) (R_\Delta) \,\, = \,\, 
M^{(0)}_{\BH} - M^{(n)}_{\BH} \,\, = \,\, \int_0^{R_\Delta}\, 
(\kappa_{(0)}-\kappa_{(n)}) (r) \d r$$
and ii) ensures that the integrand is positive.  

Next, we can put a lower bound on $\beta_{(n)}$.  We will use the fact
that, for any fixed $n$ the ADM mass $M^{(n)}_{\ADM}(R_\Delta)$ is a
monotonically increasing function of $R_\Delta$.  Since
$M^{(n)}_{\sol}$ is fixed on the $n$-th branch of static solutions, it
follows from (\ref{ymmass}) that

\medskip\noindent v) \textit{$M^{(n)}_{\BH}(R_\Delta)$ is also a
monotonically increasing function of $R_\Delta$.  (\ref{hormass}) now
implies that, for all $n$ and $R_\Delta$, $\beta_{(n)}$ and
$\kappa_{(n)}$ are positive functions of $R_\Delta$.  Hence
$M^{(n)}_\Delta$ is also positive for all $n$ and $R_\Delta$, except
for $R_\Delta =0$ where it vanishes.}

\medskip
\noindent Finally, the model has a prediction on the asymptotic
behavior (for large $R_\Delta$) of $M^{(n)}_{\BH}(R_\Delta)$,
$\beta_{(n)}(R_\Delta)$ and $\kappa_{(n)}(R_\Delta)$.  Using the known
property of the ADM mass, $M^{(n)}_{\ADM}(R_\Delta) >
M^{(0)}_{\ADM}(R_\Delta) = {R_\Delta/ 2G}$, we conclude:
$$M^{(n)}_{\BH}(R_\Delta)\,\, = \,\, M^{(n)}_{\ADM} (R_\Delta) 
- M^{(n)}_{\sol}
\,\, > \,\, {R_\Delta \over 2G} - M^{(n)}_{\sol}\, .$$
Hence, using prediction i) we conclude that, for any given $n$, the
curve $M^{(n)}_{\BH}(R_\Delta)$ lies between the two parallel lines
$f_1(R_\Delta) = {R_\Delta/ 2G}- M^{(n)}_{\sol}$ and $f_2(R_\Delta) =
{R_\Delta/ 2G}$.  Furthermore, since $M^{(n)}_{\BH}(R_\Delta)$ is a
monotonically increasing function of $R_\Delta$, it follows that:

\medskip\noindent vi) \textit{For any $n$, the curve showing the
dependence of $M^{(n)}_{\BH}(R_\Delta)$ on $R_\Delta$ becomes
asymptotically parallel to the two lines which bound it, i.e., has
slope $1/2G$ for large $R_\Delta$.  Hence, by (\ref{hormass}), the
curves showing the dependence of $\beta_{(n)}$ on $R_\Delta$
asymptotically approach the curve $\beta_{(0)}(R_\Delta) =1$ and the
curves showing the functional dependence of $\kappa_{(n)}$ of
$R_\Delta$ asymptotically approach the curve $\kappa_{(0)}(R_\Delta) =
{1/ 2R_\Delta}$.}
\medskip

These six properties account for all qualitative features of
Fig.~\ref{fig2} and Fig.~\ref{fig3} i.e., the dependence of
$M^{(n)}_{\BH}$ and $\beta_{(n)}$ on $n$ and $R_\Delta$.  In
particular for Fig.~\ref{fig3}, we have shown that our model implies
that the $\beta_{(n)}$ curves: a) intersect the y-axis at distinct
points between $0$ and $1$ and lie in the upper right hand quadrant,
b) the curves never intersect, c) have the property that higher the
$n$, lower the curve, and d) for large $R_\Delta$ become
asymptotically tangential to the curve $\beta_{(0)}(R_\Delta) = 1$.
To arrive at this qualitative understanding of the horizon properties,
in addition to the model, we used only two known qualitative
properties of the ADM mass \cite{review}: $M^{(n)}_{\sol}$ increases
with $n$ and, for any given $n$, $M^{(n)}_{\ADM}$ increases
monotonically with $R_\Delta$.

\subsection{Instability of colored black holes}
\label{s3.3}

We will now argue that the model also `explains' the instability of
the colored black holes in terms of the instability of the solitons.
Under small perturbations, the solitons are unstable; the energy
`stored' in the bound state of the soliton is radiated away to future
null infinity, ${ \I}$ \cite{uns2}.  In our model, each static black
hole with $n\not= 0$ is accompanied by a soliton which `hovers around
all the way to $i^+$' and carries information about its hair.
Therefore, it is natural to expect that under small perturbations in
the exterior region, this soliton would be unstable and radiate away
across ${\cal I}^+$ and $\Delta$.  At the end of the process, there
should be no solitonic residue at $i^+$.  The end product should thus
be an uncolored (i.e. Schwarzschild) black hole whose horizon mass
$M^{(0)}_{\BH}$ equals the future limit $M_{i^+}$ of the Bondi mass
along ${\cal I}^+$.  Our suggestion that the instability of the
colored black hole is due to the instability of the accompanying
soliton is supported by the fact \cite{uns} that all the colored black
holes on the $n$th branch have the same number (namely, $2n$) of
unstable modes as the $n$th soliton.  Of course the detailed features
of these unstable modes will differ especially because they are
subject to different boundary conditions in the two cases.

In general, however, even if one component of a bound system is
unstable, the total system may still be stable if the binding energy
is sufficiently large.  An obvious example is provided by the deuteron
nucleus ---a bound state of a proton and a neutron.  Therefore, to see
if our bound state of a bare black hole and a soliton is in fact
unstable, let us examine the energetics.  If the decay suggested above
is to occur, the total energy in the initial bound state, $E_\i =
M^{(n)}_{\BH} (R_\Delta^\i) + M^{(n)}_\sol$, should exceed the energy
in the final black hole $E_\f = M^{(0)}_{\BH} (R_\Delta^\f)$.  Since
the area of the final uncolored black hole is at least as large as the
area of the initial colored black hole, the decay can occur only if
the following inequality holds:
\be M^{(n)}_{\BH}  (R_\Delta^\i) + M^{(n)}_\sol > M^{(0)}_{\BH} 
(R_\Delta^\i) \, .\ee
The inequality is in fact satisfied.  For, the left side is the ADM
mass $M_{\ADM}^{(n)}(R_\Delta)$ while the right side is the ADM mass
$M^{(0)}_{\ADM}(R_\Delta)$ of the uncolored black hole of the same radius
$R_\Delta^\i$, and the ADM mass is known to increase with $n$.  Thus,
the energetics are such that the instability of the soliton can
persist also in the bound state, forcing the solitonic residue to
radiate away.

\begin{figure}
\centerline{
\hbox{\psfig{figure=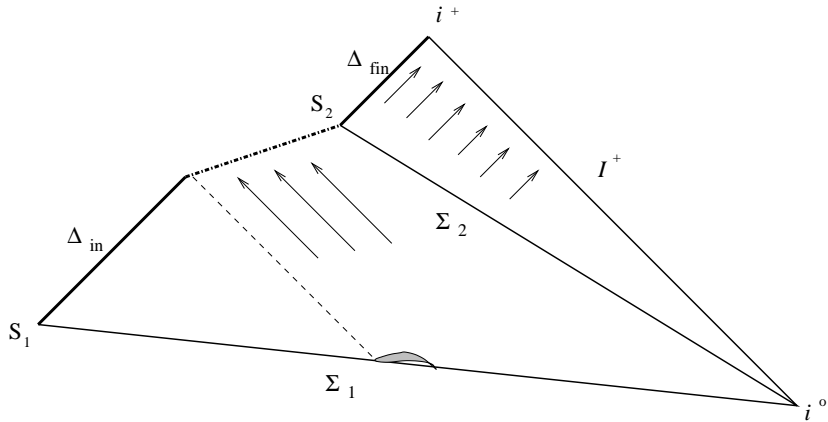,height=6cm}}}
\bigskip
\caption{An initially static colored black hole with horizon
$\Delta_{\rm in}$ is slightly perturbed and decays to a
Schwarzschild-like isolated horizon $\Delta_{\rm fin}$, with radiation 
going out to future null infinity ${\cal I}^+$.}
\label{fig4}
\end{figure}

Let us therefore assume that the process illustrated in
Fig.~\ref{fig4} does take place and examine its consequences.  Thus,
we assume that the event horizon has an initial isolated component
$\Delta_\i$ with color $n$ and radius $R_\Delta^\i$ and there is a small
perturbation on the initial partial Cauchy surface $M$.  The
perturbation evolves and, because of the instability of the soliton,
the solitonic residue is radiated away and we are left with a final,
uncolored isolated horizon $\Delta_\f$ with radius $R_\Delta^\f$
(satisfying $R_\Delta^\f \ge R_\Delta^\i$) and energy $E_{\I}$
radiated across future null infinity $\I$.%
\footnote{In a realistic situation, in distant future the event
horizon will become isolated only asymptotically and will probably not
admit a \textit{finite} isolated piece $\Delta_\f$.  However, our
conclusions will continue to hold in this case if the approach to
isolation is sufficiently fast.}
The ADM mass of the space-time does not change in this process.  Hence 
neglecting the energy in the initial perturbation (i.e., in the limit
`from above' of a sequence of perturbations with decreasing energy) 
we have:
\be M^{(n)}_{\ADM} = M_{\BH}^{(n)}(R_\Delta^\i) + M^{(n)}_{\sol} =
M^{(0)}_{\BH} (R_\Delta^\f) + E_{\I}\, . \ee
Using (\ref{binding}) and the expression of the ADM mass of the uncolored 
(i.e. Schwarzschild) black hole in terms of its horizon radius, this 
equality can be re-expressed as 
\be \label{cons}
\left[ M^{(n)}_\sol - \mid E_\b^\i\mid \right] = {1\over 2G}
\left(R_\Delta^\f -R_\Delta^\i\right) + E_\I  \, .
\ee 
Thus, the `available energy' for the process is given by:
\be \label{avail}
E_\a^{(n)}= M_\sol^{(n)}  - \mid E_\b^\i\mid\, . \ee 
Note that $E_\a^{(n)}$ can be computed \textit{knowing just the
initial configuration}.  However, the distribution of this energy in
to a part which goes in to increasing the horizon area and the part
which gets radiated to null infinity will depend on the details of the
initial perturbation.

Our model enables us to make further qualitative predictions about
this process.  Note first that, since for a fixed $n$, the binding
energy $\mid E_\b^\i\mid $ is a monotonically increasing function of
$R_\Delta^\i$ and $M^{(n)}_{\sol}$ is independent of $R_\Delta^{(n)}$,
the available energy $ E^{(n)}_\a$ decreases as $R_\Delta^\i$
increases.  In this sense, for a given $n$, a larger colored black
hole is less unstable than a smaller one.  This prediction is
supported by the fact that for the $n=1$ colored black holes, the
frequency of all unstable modes is a decreasing function of the area,
whence the characteristic decay time grows with area 
\cite{review,Bizon:2000ew}.
Next, let us fix $R_\Delta^{\i}$ and let $n$ vary.  Then, since the
available energy can be re-expressed as $E^{(n)}_\a= M_{\ADM}^{(n)} -
M_{\ADM}^{(0)}$, it follows that it increases with $n$.  In this
sense, for a given $R_\Delta^\i$, higher the $n$, more unstable the
colored black hole.  This prediction is supported by the fact that the
number of unstable modes is given by $2n$.  It would be interesting to
analyze if there exist further, more detailed correlations.

Finally, it is natural to ask if the horizon radius necessarily grows 
in this process.  Using an additional input, we will now show that the 
answer is in the affirmative.  Recall from \cite{ac:ds,afk} that each
isolated horizon in the Einstein-Yang-Mills theory carries a 
well-defined magnetic charge $P_\Delta$:
\be \label{magnetic}
P_\Delta = - {1 \over 4\pi} \,\,\oint \mid F\mid \,\, \epsilon \ee
where the integral is taken over any 2-sphere cross-section of
$\Delta$, $\epsilon$ is the area 2-form on this cross-section, and
$\mid F\mid = [ {\rm Tr} (F_{ab}\epsilon^{ab})\,\,
(F_{cd}\epsilon^{cd}) ]^{\textstyle{1\over 2}}$.  The additional fact
we need is that $P_\Delta$ vanishes if and \textit{only} if $n =0$.
Therefore, in the process under consideration, $P_\Delta$ necessarily
changes.  Let us now \textit{assume} $R_\Delta^\i = R_\Delta^\f$ and
arrive at a contradiction.  Since the expansion of the null normal
$\ell^a$ of any event horizon is necessarily non-negative, it follows
that the expansion is necessarily zero.  The Raychaudhuri equation
then implies $T_{ab}\ell^a v^b =0$ for all vectors $v^a$ tangential to
the event horizon and the form of the Yang-Mills stress-energy
$T_{ab}$ then implies that the field strength must satisfy $F_{ab}
\ell^a v^b =0$.  One can now use the reasoning of \cite{afk} (see Eq
(VI.7) to Eq(VI.9) of that paper) to conclude that $P_\Delta$ is
conserved all along the event horizon (even though some of the
isolated horizon boundary conditions may be violated in the
intermediate region).  This is however impossible because while
$P_\Delta^\i$ is non-zero $P_\Delta^\f$ vanishes because the end
product of the process is a Schwarzschild black hole.  Hence our
assumption $R_\Delta^\i = R_\Delta^\f$ can not hold; the area of the
event horizon must \textit{necessarily increase} in this
process. Hence (\ref{cons}) implies:
\be E_\I = \left( M^{(n)}_\sol - \mid E_\b^\i\mid \right) - {1\over
2G} \left(R_\Delta^\f -R_\Delta^\i\right) < E^{(n)}_\a \, .  \ee
There may well exist a more sophisticated versions of this argument
which provide stronger bounds on the change in the horizon area in
terms of $P_\Delta^\i$ and $R_\Delta^\i$ and hence a better upper
bound on the energy that can be radiated away to $\I$.

\subsection{Instability of Embedded Abelian solutions}
\label{s3.4}

Let us now consider the `embedded Abelian solutions' with magnetic
charge. In the Einstein-Maxwell theory, the electromagnetic connection
is defined on a non-trivial $U(1)$ bundle of a fixed Chern-class,
corresponding to the fixed value $P_0$ of the magnetic charge while in
the non-Abelian case, the Yang-Mills potential is defined on a trivial
$SU(2)$ bundle. (In the non-Abelian case, $P_0$ has the same
dimensions as the inverse $1/g$ of Yang-Mills coupling constant. In
the conventions/units generally used in the literature, $P_0 = \pm 1$.)
In either case, a detailed analysis shows that Eqs (1) and (2)
continue to hold, without any extra terms.  Since the presence of an
electric charge plays no essential role in this analysis, for
simplicity, let us focus on the sector of the phase space where it
vanishes. Then, in both cases, we are left with a 1-parameter family
of static solutions which can be coordinatized by the value of the
horizon area. Using the expression of the surface gravity
$$\kappa_{(P_0)} = {1 \over 2R_\Delta}\left( 1 - 
{GP_0^2\over R_\Delta^2}\right)$$ 
of this family, we can introduce a permissible evolution field $t_0$
and obtain the corresponding first law:
$$ \delta E^{(t_0)}_\Delta = \frac{1}{8\pi G}\kappa_{(P_0)} \delta
a_\Delta $$
Let us integrate this law to obtain the horizon Energy:
\be \label{C} E_\Delta^{(t_0)} (R_\Delta) = {R_\Delta \over 2G }
\left( 1 + {GP_0^2 \over R_\Delta^2}\right) +C \ee
where $C$ is the integration constant. For colored black holes, the
constant was determined \cite{ac:ds,afk} by requiring that
$E_\Delta^{(t_0)}$ should vanish in the limit $R_\Delta$ goes to
zero. For (embedded) Abelian black holes under consideration, by
contrast, the minimum value of $R_\Delta$ is $\sqrt{G}\mid P_0\mid$
when the black hole becomes extremal.  Therefore, we need a new
strategy to determine the constant $C$.

In the Einstein-Maxwell case, it is natural to appeal to the duality
invariance of the theory.  The sector of the theory with zero magnetic
charge but arbitrary electric charge $Q$, we know \cite{abf,afk} that
the horizon energy is given by%
\footnote{In the reasoning used in \cite{abf,afk}, a key role is
played by the fact $Q$ is allowed to vary, i.e., that the first law
reads: $\delta M_\Delta = (\kappa/8\pi G) \delta a_\Delta + \Phi
\delta Q$, and by the fact that on this phase space with variable $Q$,
there is no constant of dimensions of energy we can construct from the
only available constants $G$ and $c$.}:
\be E_\Delta^{(t_0)} (R_\Delta) = {R_\Delta \over 2G }
\left( 1 + {GQ^2 \over R_\Delta^2} \right) \ee
Hence, by duality invariance, we conclude $C=0$.%

One could also have reached the same conclusion by starting ab-initio
in the dual description, i.e., by considering from the beginning the
vector potential for the electric field as the configuration variable
and the magnetic field as the momentum variable and allowing the
magnetic charge $P$ to be arbitrary but setting the electric charge to
zero from the beginning.  One could then work on a trivial bundle and,
using the dual of the standard framework of \cite{afk}, conclude that
$C =0$.  To summarize, because of our choice of the evolution vector
field $t_0^a$, we can interpret $E^{(t_0)}_\Delta$ as the horizon
mass $M_{\BH}^{\rm EM}$ of the Einstein-Maxwell magnetically charged
black holes and each of the two arguments given above leads us to the
result
\be \label{emaxmass} 
M_\BH^{\rm EM} = {R_\Delta \over 2G}\left( 1 + {GP_0^2 \over 
R_\Delta^2}\right) \, .  \ee

Let us now turn to Yang-Mills theory.  To determine the constant $C$
in (\ref{C}), we can no longer appeal to duality invariance or use a
dual description with the vector potential of the electric field as
the configuration variable.  Thus, we need a new input. Let us first
consider the colored, spherically symmetric, \textit{non-Abelian}
black hole space-times discussed in the last sub-section.  It is known
that as $n$ tends to zero, these colored, \textit{non-Abelian} black
hole space-times tend to the magnetically charged, embedded Abelian
black hole space-time provided the horizon radius satisfies the bound
$R_\Delta \ge \sqrt{G}P_o$ \cite{review}.  If $R_\Delta = \sqrt{G}\mid
{P_0}\mid$, the magnetically charged black hole is extreme.  The new
input is that as $n$ tends to infinity, $\beta_{(n)}(R_\Delta)$ tends
to zero for $0\le R_\Delta \le \sqrt{G}P_0$ (see Fig.~\ref{fig3}).
Hence, using (\ref{hormass}) it follows that as $n$ tends to infinity,
keeping $R_\Delta = \sqrt{G}P_0$, the horizon mass
$M^{(n)}_{\BH}(R_\Delta)$ tends to zero.  Hence, in the Yang-Mills
phase space, we are led to set the constant $C$ of (\ref{C}) to: $C =
- P_0/\sqrt{G}$.  Consequently, in Yang-Mills theory, the horizon mass
of the embedded Abelian solutions is given by \cite{ac:ds}:
\be \label{eymmass}
M_{\BH}^{\rm EYM} = {R_\Delta \over 2G}\left( 1 + {GP_0^2 \over 
R_\Delta^2}\right) - {\mid P_0\mid \over \sqrt{G}} \, .  \ee
Thus, the horizon mass in the Yang-Mills theory is \textit{lower} than 
that in the Einstein theory.

It is this difference that accounts for the strikingly different
stability properties of these black holes in the two theories. Since
the solutions are asymptotically flat at spatial infinity in the
standard sense, their ADM mass is determined only by the asymptotic
behavior of the space-time metric and is given by:
\be
M_\ADM =  {R_\Delta \over 2G}\left( 1 + {GP_0^2 \over R_\Delta^2}
\right)
\ee
in \textit{both} theories. Thus we conclude:
\be 
M_\ADM - M_\BH^{\rm EM} = 0, \quad {\rm while} \quad
M_\ADM - M_\BH^{\rm EYM} = {\mid P_0\mid \over \sqrt{G}}
\ee
Thus, as guaranteed by general arguments from symplectic geometry, the
difference between the ADM and the horizon mass is indeed constant on
the connected component of magnetically charged static solutions but
the value of the constant is theory dependent. In the Einstein-Maxwell
theory, the value is zero. All the energy in the solution is
associated with the horizon and there is no provision in the `energy
budget' for energy to be radiated away. In the Einstein-Yang-Mills
theory, the energy budget does have such a provision. The future limit
of the horizon mass ($= M_\BH^{\rm EYM})$ is \textit{less} than the
future limit of the Bondi mass along $\I$ ($= M_\ADM$). Therefore,
there is a mismatch at $i^+$. Although in the limit $n \to \infty$
there is no regular soliton,\footnote{It has been shown that for 
$r >1$ the limiting solution approximates the extremal RN solution
\cite{Breitenlohner:1995qc}}
since
$$ \lim_{n\to \infty} M^{(n)}_\sol \not= 0 \, ,$$
from the standpoint of energetics, there is still a solitonic residue
at $i^+$. Hence, as a limiting case of the situation considered in the
last sub-section, our reasoning implies that the embedded Abelian
black holes should be unstable in the Einstein-Yang-Mills theory even
though their Einstein-Maxwell counterparts are stable. Detailed
analysis \cite{stability} has shown that this is precisely what
happens! Furthermore, we can compute the `available energy' for the
black hole of radius $R_\Delta$ in EYM. It has the form $E_{\rm
avail}^{\rm EYM}=P_0^2/2R_{\Delta}$. This suggests that the
frequencies of the unstable modes will be smaller for larger black
holes.

It may seem surprising at first that there is such a marked difference
in the stability properties in spite of the fact that the space-time
geometries of these Einstein-Maxwell and Einstein-Yang-Mills black
holes are identical. However, since these geometries `live' in
distinct phase spaces, modes which can trigger instability within one
phase space can be absent in the other. Indeed, stability is not a
property of a specific solution in isolation but refers also to other
`accessible' states and the set of accessible states is different in
the two phase spaces. The notion of the horizon mass can capture this
difference because it is a \textit{phase space} notion rather than a 
space-time notion.

To summarize, these magnetically charged black holes bring out the
fact that the horizon mass $M_\BH$ of an isolated horizon $\Delta$ can
not be determined knowing only the space-time geometry unless one
knows before hand which phase space this space-time belongs to. At
first, this `ambiguity' in the horizon mass seems surprising. However,
as we have seen, it has a deep physical implication. In the next
section, we will see that a further subtlety associated with the
notion of horizon mass arises as we consider theories with additional
dimensionful constants and more complicated hair.
 
\section{Hairy black holes in more general theories}
\label{s4}

\begin{figure}
\centerline{
\hbox{\psfig{figure=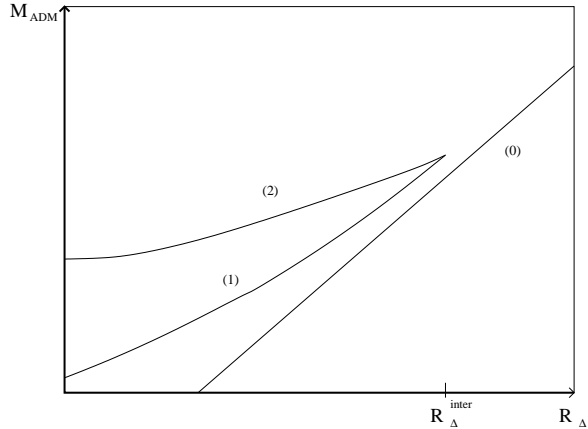,height=6cm}}}
\bigskip
\caption{The ADM Mass as a function of the horizon radius $R_\Delta$
in theories with a built-in non-gravitational length scale. The
schematic plot shows crossing of families labelled by $n=1$ and $n=2$
at $R_\Delta = R^{\rm inter}_{\Delta}$. }\label{fig5}
\end{figure}

Let us now turn to more general theories with hair, by allowing Higgs
or Procca fields in addition to the Yang-Mills fields, or considering
Einstein-Skyrme theories \cite{pb:tc,tt:km,review}.  These theories
have additional dimensionful constants which trigger new phenomena.
The static black hole sector of such theories enables one to analyze
these novel aspects of the interplay between gravity and the nonlinear
matter fields. Of primary interest to us is the `crossing phenomena'
of Fig.~\ref{fig5} where curves in the `phase diagram' (i.e. plot of
the ADM mass versus horizon radius) corresponding to the two families
of static solutions intersect. With due attention to this phenomenon,
considerations of Section \ref{s3} will extend to hairy black holes in
these more general theories. In this section we will discuss the
relevant subtleties that must be taken in to account.

We will begin with a qualitative argument to elucidate the `origin' of
the crossing phenomenon.  Let us first ignore gravity and consider
these theories in Minkowski space-time. The coupling constants and
other parameters (such as the vacuum values of fields) in these
theories provide a length scale which determines the `size' $R^{\rm
Mink}_{\sol}$ of the solitonic solution. Let us now couple these
theories to general relativity. Gravitational effects, being
attractive, can only reduce this size. Hence, if a black hole is to
exhibit any hair at all, its horizon radius $R_\Delta$ must be
\textit{less than} $R^{\rm Mink}_{\sol}$.  (In fact, using more
detailed arguments involving the pressure of matter fields near the
horizon and at infinity, one can give a sharper bound: $R_\Delta <
(2/3) R^{\rm Mink}_{\sol}$ \cite{nqs}.) Thus, in these theories which
have an in-built length scale already in Minkowski space-time, the
horizon radius $R_\Delta$ of hairy black holes is bounded above by
that length scale which is independent of Newton's constant. Note
that, Yang-Mills theory in Minkowski space has no such built-in length
scale whence there is no bound on the horizon radius of hairy, static
Einstein-Yang-Mills black holes.

If there is an upper bound to the horizon radius, it is natural to ask
what happens in the phase diagram to the individual, connected
branches corresponding to static black holes. Recall first that we can
view the space of static black hole solutions as the manifold of
extrema of the ADM mass functional on phase space \cite{ds:rw}.  In
the Einstein-Yang-Mills theory, each connected component of this
manifold is a sub-manifold without boundary. Let us suppose that the
topology of these sub-manifolds does not change abruptly as the new
coupling (and other) constants are slowly switched on to arrive at
more general theories.  Then, if the horizon radius of the new
theories is bounded from above, the sub-manifolds which are
disconnected in the Einstein-Yang-Mills theory must merge smoothly in
these theories.%
\footnote{In the case of the `embedded Abelian family', by contrast,
the sub-manifold of static solutions has a boundary corresponding to
extremal black holes. However, in this case, the phase space itself
has a boundary consisting of space-times whose isolated horizons have
zero surface gravity and the sub-manifold just ends at this boundary.
If sub-manifolds of static \textit{colored} black holes were to
terminate without merging, these sub-manifolds would have a true
boundary in the \textit{interior} of the phase space.}
Then, in the projection of this smooth sub-manifold to the $(R_\Delta,
M_{\ADM})$ plane, one would see two branches which cross at the point
corresponding to the maximum value of $R_\Delta$. This is indeed what
is found in numerical investigations of the spherically symmetric,
static black hole sectors of Einstein-Skyrme,
Einstein-Yang-Mills--Proca, and Einstein-Yang-Mills-Higgs theories
\cite{pb:tc,tt:km,torii} and, more recently, in the static axially
symmetric black hole sector of Einstein-Yang-Mills-Higgs theory
\cite{bk:jk2}. Our discussion suggests that this crossing phenomenon
will occur in any theory that contains a length scale independent of
$G$ and allows hairy black holes.

Let us start by analyzing the way two branches merge at the cross-over
point. For simplicity, let us focus on static black holes that carry
no other charge \textit{at infinity} beyond the ADM mass. As long as
the energy momentum tensor in the theory satisfies appropriate energy
conditions, the static black holes can be assumed to admit a Killing
horizon with a bifurcation two surface \cite{ir:rw}.  For such black
holes, arbitrary perturbations are known to obey the first law of
black hole dynamics \cite{ds:rw}:
\be \label{1law:adm}
\delta M_{\ADM} =\frac{1}{8\pi}\kappa\; \delta a_{\Delta}\, . \ee 
Now consider the phase diagram (i.e., the plot of the ADM mass as a
function of the horizon radius) for this family and focus on the
intersection point between the two branches. If the two curves
intersect at a finite angle, the first order change in the mass for a
given infinitesimal change in the horizon area would depend on the
branch used to approach the intersection point. In view of
(\ref{1law:adm}), this would contradict the fact that the surface
gravity has a well defined value at the solution corresponding to the
intersection point. Hence, we conclude:
\medskip

\noindent \textit{vi)' at any cross-over point, the two branches must
have the same tangent vector.}
\medskip

\noindent Explicit calculations have confirmed this behavior in all
theories where crossing phenomena has been seen to occur. 

Next, let us examine how the crossing phenomenon affects the notion of
the horizon mass. As in Section \ref{s3.1}, for all theories under
consideration, given any permissible evolution vector field $t^a$ we
are again led to a horizon energy $E^{(t)}_\Delta$ and the
corresponding first law (\ref{1law}). The delicate point again is the
selection of a preferred evolution vector field. Because of the
presence of hair ---more precisely, because the value of the horizon
area fails to determine a static black hole uniquely--- we can no
longer use the strategy that was successful in the Einstein-Maxwell
case to define a canonical evolution vector field $t_0^a$ and use
$E^{(t_0)}_\Delta$ as the horizon mass $M_\Delta$ on the \textit{full}
phase space. However, as in the Einstein-Yang-Mills case, we can again
focus on any one branch, say the $n$th, of static solutions, use the
surface gravity $\kappa_{(n)}(R_\Delta)$ to select a permissible
evolution vector field $t^a_{(n)}$ and construct the horizon energy
$E^{(n)}_\Delta$ and set (a la (\ref{hormass})) the mass $M^{(n)}_\BH$
of any black hole on the $n$th branch to be:
\be\label{hormass2} 
M^{(n)}_{\BH}(R_\Delta) = E^{(n)}_\Delta (R_\Delta)\, = 
\frac{1}{2G}\int_0^{R_\Delta}
\beta_{(n)}(r) \d r\, .  \ee
This strategy is tenable except at the cross-over points. At these
points, there is a problem: Because the function $\beta_{(n)}$ depends
on $n$, the value of the horizon mass is now ambiguous. If branches
$n$ and $n+1$ cross, as in Section \ref{s3.2} we have
$\beta_{(n+1)}(R_\Delta) < \beta_{(n)}(R_\Delta)$, whence
$M^{(n+1)}_\BH < M^{(n)}_\BH$. We saw in Section \ref{s3.4} that,
since the notion of the horizon mass refers to the phase space, the
value of the horizon mass of a black hole is not determined simply by
its space-time geometry. Here we see another, and starker, aspect of
the same phenomenon.

In the magnetically charged black holes of Section \ref{s3.4}, the
same space-time appeared in two different phase spaces and the value
of its horizon mass depended on which phase space one used. In the
present case, there is a single phase space. However, we can use
either the $n$th family of static solutions and choose a permissible
evolution vector field $t^a_{(n)}$ using the functional dependence of
the surface gravity $\kappa_{(n)}$ on the horizon radius $R_\Delta$ to
fix the normalization, or we can do the same using the $(n+1)$th
family. Since the functional dependences of $\kappa_{(n)}$ and
$\kappa_{(n+1)}$ on the horizon radius are necessarily different, we
obtain two distinct vector fields and two distinct Hamiltonians which
generate time-evolution along them. Each of the resulting horizon
energies $E^{(n)}_\Delta$ and $E^{(n+1)}_\Delta$ is a well-defined
function on \textit{the entire phase space} and leads to a first law
(\ref{1law}). Although the two energies differ, there is complete
consistency because the two energies arise from two different
evolutions (i.e., two different normalizations of the evolution vector
field on the horizon). The ambiguity arises only when we want to
identify these energies with horizon masses and is confined just to
the cross-over points. As remarked in the Introduction, definitions of
quasi-local mass in general relativity invariably encounter unexpected
features and the ambiguity in the horizon mass that arose (in Section
\ref{s3.4} and) here is the counter-intuitive feature encountered in
the present Hamiltonian approach.

However, in spite of ---or rather, because of--- this ambiguity, one
can extract useful information from this definition of horizon mass.
Let the $n$th and $(n+1)$th branch intersect at the horizon radius
$R^{\in}_\Delta$. Then, using (\ref{ymmass}) it follows that:
\be
M^{(n)}_{\ADM}(R_\Delta)- M^{(n)}_{\BH}(R_\Delta)
=M^{(n)}_{\sol}\quad {\rm and} \quad
M^{(n+1)}_{\ADM}(R_\Delta)- M^{(n+1)}_{\BH}(R_\Delta)
=M^{(n+1)}_{\sol}\, . 
\ee
Hence, using (\ref{hormass}), at the intersection point we obtain:
\be
M_{\ADM}(R^\in_\Delta) =M^{(n)}_{\sol} +
\frac{1}{2G}\int_0^{R_{\Delta}^\in} \,\beta_{(n)}({r})\, \d r
= M^{(n+1)}_{\sol} +
\frac{1}{2G}\int_0^{R_{\Delta}^\in} \,\beta_{(n+1)}({r})\, \d r
\ee
The second equality yields: 
\be M^{(n+1)}_{\sol}-M^{(n)}_{\sol}=
\frac{1}{2G}\int_0^{R_{\Delta}^\in} \,\beta_{(n)}({r})\, \d r
-\frac{1}{2G}\int_0^{R_{\Delta}^\in} \,\beta_{(n+1)}({r})\, \d r =
\frac{1}{2G} \oint \beta (r)\, \d r \label{26}
\ee 
where the last closed counter integral is performed by first moving
along the $n$th branch from $R_\Delta = 0$ to $R_\Delta =
R_\Delta^\in$, then moving back along the $n+1$th branch to $R_\Delta
=0$ and finally sliding down the y axis to the point $R_\Delta = 0$ on
the $n$th branch (see Fig.\ref{fig5}).%
\footnote{Since the `y-axis' corresponds to zero area, points on it do
not belong to the black hole phase space considered here.
Therefore, one should close the contour by moving along the vertical
line $R_\Delta =\epsilon$ and then take the limit $\epsilon \to 0$.} 
Note that the counter integral is non-zero because $\beta_{(n)}$ and
$\beta_{(n+1)}$ are \textit{distinct functions} of $R_\Delta$, or
equivalently, \textit{precisely} because the horizon mass undergoes a
jump at the cross-over point.  This equation is striking in that it
provides a quantitative relation between the horizon properties of
hairy black holes and masses of solitons, even though the two belong
to completely different sectors of the theory. The equality was first
checked numerically in certain static, axi-symmetric hairy black hole
solutions in the Einstein-Yang-Mills-Higgs theory\cite{bk:jk2}.  Given
the generality of our considerations, one expects this relation to
hold also for more general matter sources which lead to the crossing
phenomena, such as Yang-Mills-Higgs, Yang-Mills-Proca and Skyrme
fields.

Finally, our discussion of the qualitative behavior of equilibrium
properties (Section \ref{s3.2}) and instability of colored
Einstein-Yang-Mills black holes (Section \ref{s3.3}) will extend also
to these more general black holes with two provisos: a) due care is
taken of the ambiguity in $M_{\BH}$ at the cross-over points; and, b)
property vi) in Section \ref{s3.2} regarding the asymptotic behavior
of horizon quantities for large $R_\Delta^{(n)}$ replaced by vi)' noted
above on the behavior of the phase diagrams at the cross-over points.

\textit{Remark:} The above discussion can be used also to shed light
on the behavior of the ADM mass of \textit{solitons} in the
\textit{Einstein-Yang-Mills theory}, in particular its dependence on
$n$.  Let us begin with the collection of Einstein-Yang-Mills-Higgs
theories parametrized by the vacuum expectation value $\eta$ of the
Higgs field. Now, the static sector of these theories consists of
extrema of the ADM mass at a fixed area of the internal boundary
\cite{ds:rw} and, when $\eta =0$, the extrema are configurations where
the Higgs field vanishes identically. Thus, we can consider the static
sector of the Einstein-Yang-Mills theory as the limit $\eta \to 0$ of
the static sector of the Einstein-Yang-Mills-Higgs theory. Now, in the
Einstein-Yang-Mills-Higgs theories, the crossing phenomenon occurs and
the masses of solitons are related by Eq. (\ref{26}) for all values of
$\eta$. Taking the limit when $\eta \to 0$ we arrive at an identical
equation, now relating the ADM masses of the Einstein-Yang-Mills
solitons.  In this limiting case, the branches do not cross at a
finite value of the horizon radius. However, from the discussion at
the beginning of this section, one expects that as $\eta$ tends to
zero, the value of the horizon radius at the crossing point tends to
infinity. Now, from the Fig.~\ref{fig3}, and more generally from
prediction vi), we see that $(\beta^{(n)} - \beta^{(n+1)})(r) \to 0$
as $R_\Delta \to \infty$. Thus, in the limit, all the $\beta$ curves
meet at infinity, including the Schwarzschild branch corresponding to
$\beta_{(0)}(R_\Delta) =1$.  Hence, it is plausible that the area
bounded by the two curves $\beta_{(n)}$ and $\beta_{(n+1)}$ in the
plot of $\beta$ as a function of $R_\Delta$ is finite, i.e., that the
contour integral on the right side of (\ref{26}) is well-defined also
in the Einstein-Yang-Mills theory. Thus, we obtain an explicit
prediction for the value of the ADM mass of the $n$-th soliton in this
theory:
\be \label{27}
 M^{(n)}_{\sol} =\frac{1}{2G}\int_0^\infty  (1-\beta_{(n)} (r))\, \d r\, . 
\ee
Again, the equation is striking because it enables one to compute
\textit{soliton} masses from surface gravity of \textit{black holes},
now in the Einstein-Yang-Mills theory.  To our knowledge, this
prediction has not been tested. Note, however, that it uses not only
the framework presented in this paper, but also a continuity
assumption on the embedding of the static sector of the
Einstein-Yang-Mills theory in that of the
Einstein-Yang-Mills-Higgs theories.
Nonetheless, a qualitative
consequence of (\ref{27}) is borne out by numerical simulations.  The
behavior of $\beta_{(n)}$ in Fig.~\ref{fig3} suggests that 
$M^{(n+1)}_{\sol} - M^{(n)}_{\sol}$ approaches zero as $n$ tends to
infinity and this behavior is supported by the numerical study of low
$n$ soliton solutions \cite{bk,review}, and by rigorous methods
in \cite{Breitenlohner:1994es}.
It would be interesting to test (\ref{27}) in a more direct ways.
Finally, note that if we combine Eq. (\ref{27}) with Eq. (\ref{ymmass})
we arrive at an expression for $M_{\ADM}(R_\Delta)$,
\be
M_{\ADM}^{(n)}(R_\Delta)=\frac{R_{\Delta}}{2G}+
\frac{1}{2G}\int_{R_\Delta}^\infty  (1-\beta_{(n)} (r))\, \d r
\ee
The first term is what we have called the `bare mass', and the second
term could be interpreted as the `mass of the hair', corresponding to the
soliton mass plus the binding energy as in Eq. (\ref{binding}).

\section{Discussion}
\label{s5}

Because of the black hole uniqueness theorems in Einstein-Maxwell
theory \cite{pc,mh} it was widely believed that black holes would have
no hair also in theories more complicated, non-Abelian gauge fields.
As soon as this expectation was shown to be incorrect \cite{pb1,pb2},
the subject of hairy black holes became a focal point in the
mathematical physics literature. By now, static hairy black hole
solutions have been found numerically for a variety of matter sources
coupled to general relativity. Their horizon attributes, such as the
dependence of the surface gravity on the horizon area, have been
plotted and their stability has been studied in some detail. The
literature in the field has grown steadily over the last decade with
discoveries of new families of solutions made by combining analytic
and numerical techniques. Therefore, a long list of facts regarding
hairy black holes is now available. In other branches of physics, when
sufficient data accumulates either through experiments or numerical
simulations, the subject is generally considered to be ripe for
phenomenology; simple heuristic models are used to put `order' in the
data. In a similar spirit, we introduced a heuristic model of hairy
black holes as bound states of ordinary black holes and solitons and
used it to account for qualitative features of a significant fraction
of the accumulated data.

The notion of the horizon mass played a key role both in motivating
the model and in its applications. In particular, this notion could be
used effectively to gain qualitative insight in to `why' hairy black
holes are unstable, and more strikingly, `why' the magnetically
charged Reissner-Nordstr\"om black holes are unstable in the
Einstein-Yang-Mills theory although they are stable in the
Einstein-Maxwell theory. Since horizon mass plays an important role in
our phenomenology, we also investigated certain subtleties and pointed
out some of its counter-intuitive properties. Finally although for
definiteness we restricted ourselves to the Einstein-Yang-Mills theory
in Section \ref{s3}, our phenomenological model is valid also in
theories with more complicated matter sources and our qualitative
analysis can be repeated in this more general context provided one
takes an appropriate care of the subtleties encountered in Section
\ref{s4} at the cross-over points.

\acknowledgments We would like to thank S. Fairhurst for
discussions, P. Bizon, P. Forgacs and S.T. Yau for pointing out
some references, and U. Nucamendi for help with the figures.
This work was supported in part by NSF grants
INT9722514, PHY95-14240, PHY00-90091, the Eberly research funds of
Penn State, DGAPA-UNAM grant No IN121298, CONACyT grants J32754-E and
32272-E, and by an NSF-CONACyT collaborative grant.

\end{document}